# Robust Probabilistic Analysis of Transmission Power Systems based on Equivalent Circuit Formulation

Martin R. Wagner, Amritanshu Pandey, Marko Jereminov, Larry Pileggi
Dept. of Electrical and Computer Engineering
Carnegie Mellon University
Pittsburgh, PA

*Abstract*— Recent advances in steady-state analysis of power systems have introduced the equivalent split-circuit approach and corresponding continuation methods that can reliably find the correct physical solution of large-scale power system problems. The improvement in robustness provided by these developments are the basis for improvements in other fields of power system research. Probabilistic Power Flow studies are one of the areas of impact. This paper will describe a Simple Random Sampling Monte Carlo approach for probabilistic contingency analyses of transmission line power systems. The results are compared with those from Monte Carlo simulations using a standard power flow tool. Lastly, probabilistic contingency studies on two publicly available power system cases are presented.

I. INTRODUCTION

Steady-state analysis of the power system used for planning and operation of the power grid has been traditionally performed using power flow analysis, where power systems represented by deterministic sets of load and generator macro-models are solved to obtain their steady-state operating point. The traditional approach to solve this problem is by simulating it using the PQV formulation based on highly nonlinear power mismatch equations and complex state variables represented in polar coordinates. There have been attempts to reduce the nonlinearity of the PQV formulation by representing the power flow problem using the current mismatch equations. The first Current Mismatch algorithm for modeling the PQ bus was introduced in [1], and further extended to a PV bus model in [2].

Reference [2] also comments that a breakthrough in power flow robustness would spark insight into more complex power system problems, such as power system dynamics, state estimation, contingency analysis, steady-state optimization, and others. A breakthrough in power flow simulation was recently achieved by reformulating the problem in terms of true physics based current and voltage state variables and simulating it as an equivalent circuit [3]. Similar to the current injection approach, the complex governing circuit equations are formulated based on underlying relationships between currents and voltages given by Kirchhoff's Current Law (KCL) and Kirchhoff's Voltage Law (KVL). To further enable the application of Newton Raphson for finding the operating point of the formulated complex circuit, the respective equations are split into real and imaginary parts, which correspond to the splitting of a complex circuit to real and imaginary sub-components coupled by controlled source circuit elements. Modeling the power flow problem in terms of an equivalent split circuit further allows the application of decades of circuit simulation research, where simulation techniques are well understood and proven to reliably find the correct physical solution for highly nonlinear large scale problems. Most importantly, the equivalent circuit formalism enabled the development of new simulation algorithms specifically designed to robustly solve large power flow problems. For instance, a new Tx-stepping continuation method was demonstrated to provide robust solution for systems as large as 85k+ buses [4]. This newfound robustness sparks possibilities in many other areas of power system research. One apparent area profiting from these advancements are Probabilistic Load Flow (PLF) studies, which enhance the power flow problem by taking uncertainties into account. Probabilistic methods in power systems are used to address a variety of topics ranging from long term planning [5] to study short-term operational challenges [6]. While analytic formulations for PLF exist a numerical approach using different forms of Monte Carlo simulations is a widely used alternative [7],[8] due to its generality.

Of particular interest for statistical methods is contingency analysis, which is a vital tool for system operation as well as system planning. In general, contingency scenarios are studied in a deterministic manner, limiting the possible outcomes to feasible/infeasible statements created by an underlying power flow software. Recent advances have been made in probabilistic evaluation of system contingencies using analytic [9] as well as sampling methods [10]. These probabilistic contingency approaches attempt to find a measure of operational risk for a certain power system problem by including contingency likelihood as well as impact on the power system and different uncertainty measures affecting the system. However, the solutions obtained from these analyses are dependent on the robustness and generality of the underlying power flow solution. A lack of which can lead to erroneous results and cause the necessity of highly conservative decision making.

In this paper we describe an approach for the probabilistic analysis of power systems that places emphasis on the robustness of the power flow solver to produce useful statistical answers. The approach is based on the equivalent split circuit formulation and associated continuation methods that are described in the next section. Section III then describes details of the proposed method. Section IV presents results starting with a comparison of the proposed method to an approach using a standard power flow solver, and case studies of two publicly available power system cases are presented. Section V provides a conclusion and future work.

## II. BACKGROUND

### A. The Equivalent Circuit Formulation

A brief overview of the equivalent split-circuit formulation and related convergence strategies is provided in the following, starting with a description of the nonlinear load and generator models.

*1) Load Model:*

The most common model of power system loads are the PQ loads, which are defined by constant real and reactive powers. Equations (1) and (2) show how to express this model in terms of real and imaginary currents flowing into the load as function of the bus complex voltage $V = V_{RL} + jV_{IL}$.

$$I_{RL} = \frac{P_L V_{RL} + Q_L V_{IL}}{V_{RL}^2 + V_{IL}^2} \quad (1)$$

$$I_{IL} = \frac{P_L V_{IL} - Q_L V_{RL}}{V_{RL}^2 + V_{IL}^2} \quad (2)$$

These equations are nonlinear and thus have to be linearized for solution via the Newton-Raphson (NR) numerical method. For linearization, a first order Taylor expansion for (1) and (2) can be written as:

$$I_{RL}^{k+1} = \frac{\partial I_{RL}}{\partial V_{RL}}|_{V_{RL}^k}(V_{RL}^{k+1} - V_{RL}^k) + \frac{\partial I_{RL}}{\partial V_{IL}}|_{V_{IL}^k}(V_{IL}^{k+1} - V_{IL}^k) + I_{RL}^k \quad (3)$$

$$I_{IL}^{k+1} = \frac{\partial I_{IL}}{\partial V_{RL}}|_{V_{RL}^k}(V_{RL}^{k+1} - V_{RL}^k) + \frac{\partial I_{IL}}{\partial V_{IL}}|_{V_{IL}^k}(V_{IL}^{k+1} - V_{IL}^k) + I_{IL}^k \quad (4)$$

Now, (3) and (4) can be directly mapped into equivalent circuit elements. The partial derivatives correspond to either conductances, if they map a voltage to its own current, or controlled current sources, if they map a different voltage to a current. Terms of the k[th] iteration are lumped together and expressed as independent current sources.

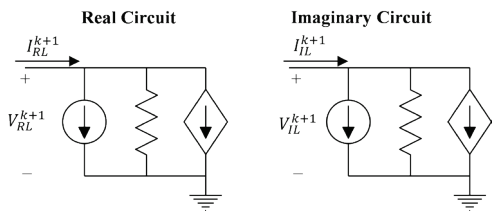

Figure 1 Linearized equivalent circuit for current mismatch equations of the PQ load model.

Figure 1 depicts the equivalent real and imaginary circuits of the PQ load model corresponding to the linearized expressions of (3) and (4).

*2) Generator Model (PV-Bus)*

A PV generator is modeled by a constant real power $P$ and a constant voltage magnitude $V_m$. The produced reactive power $Q$ is added as an additional unknown to the system of equations, requiring an additional equation for the solution of the system. The voltage magnitude constraint

$$V_m^2 = V_R^2 + V_I^2 \quad (5)$$

is included for that purpose. The remaining equations of the generator model are equal to (1) and (2) of the PQ load model, but now with Q as a state variable. Hence, (3) and (4) are adapted to include partial derivatives of Q for real and imaginary voltages for the generator model. Other power system components such as transmission lines and transformers result in linear equivalent circuit models based on similar derivations, as described in [11].

### B. Methods for robust convergence

Next we provide an overview of the convergence improvement techniques that are used to robustly solve large and ill-conditioned power flow system problems.

*Voltage Limiting* is a simple, yet powerful approach to improve convergence by limiting changes from one NR iteration to the next. In contrast to damping of the whole solution vector (damped N-R), voltage limiting only limits variables that succeed a certain threshold [3].

*Transmission Line Stepping (Tx-stepping)* was recently introduced as a specialized algorithm to improve convergence of big and ill-conditioned power flow cases [4]. It falls into the category of continuation methods. These methods aim to change the initial system in a way that a trivial solution can be found for the altered system. This trivial solution is then used as an initial guess for a successive problem that is gradually changed back towards its original form. Given a smooth path of system changes and an initial high voltage solution, these methods guarantee convergence to a correct solution of the system. Tx-stepping finds a trivial initial solution by adding high electrical admittances parallel to the branches of the system, virtually shorting each system bus to the slack bus. The initial solution for each bus is then close enough to the slack bus voltage and angle for the Newton-Raphson algorithm to converge. After this solution is found, the admittances are gradually reduced, successively leading to solutions that are closer to the original problem. In the final step, the admittances are zero and the original problem is solved.

### C. Monte Carlo Simulations

Monte Carlo (MC) simulations have proven to be effective tools in many applications ranging from finance to circuit design [7]. Convergence of the MC method is guaranteed by the probabilistic limit theory. Using random sequences created by pseudo random number generators (PRNGs) its convergence is bound by $O(1/\sqrt{N})$, where N is the number of samples.

## III. PROPOSED METHOD

Based on the robustness of the SUGAR power flow approach, a Simple Random Sampling Monte Carlo algorithm was designed to study the influence of load uncertainties on power systems and their contingencies. In the following the details of this approach will be outlined.

## A. SUGAR Monte Carlo Algorithm

Figure 2 depicts the proposed approach. The first step of the algorithm solves the power flow (PF) for the deterministic base case. The obtained solution is used as initial guess for the Monte Carlo samples. The PF solutions of the samples are computed in parallel. When a sample does not converge using only voltage limiting in the Newton Raphson solver, the algorithm falls back to solve the system using the robust Tx-stepping method. Finally, all generators of the sample are ensured to be operating inside their reactive power limits. The possible outcomes for a sample are normal power system operation, if a valid PF solution is found, voltage collapse if the robust PF algorithm did not find a feasible solution, including generator reactive power limits. If a PF solution is found but one of the operational constrains of the system is exceeded, the sample is classified as invalid, labeling the system as angular unstable, infringing on voltage bands, or exceeding branch current limits. The algorithm stops if either the maximum number of samples is reached, or the minimum confidence interval of a certain measure is fulfilled.

## B. Implementation

A C++ implementation of SUGAR is used to simulate multiple Monte Carlo samples in parallel using thread level parallelism. To maintain a low memory footprint, data structures that do not contain state of a sample are shared amongst the threads. The Simple Random Sampling Monte Carlo method relies heavily on a high quality random number sequence. In this implementation, the C++ standard library is used to create pseudo random number sequences using the 64-bit version of the Mersenne Twister 19937 PRNG.

## C. Confidence intervals

Goals of Probabilistic PF studies can be either to estimate a certain power system measure, such as voltage angle or line flow, or a property of the power system itself, such as probabilities of angular instability or voltage collapse. The two scenarios ask for different methods to find confidence intervals for the Monte Carlo simulation results. For a property that can be attributed to a binary random variable, such as the probability of a voltage collapse, the standard deviation $\hat{\sigma}_n$ is completely defined by the estimated mean probability $\hat{p}_n$. Taking the Central Limit Theorem (CLT) into account, the 99% confidence interval for such a property is

$$ci_{99} = 2.58 \sqrt{\frac{\hat{p}_n(1-\hat{p}_n)}{n}}, \qquad (6)$$

where $n$ is the number of Monte Carlo samples, the factor under the square root is the variance estimate of the CLT based normal distribution, and 99% of the values in this distribution fall in the region $\hat{p}_n \pm 2.58\,\hat{\sigma}_n$ [12].

A special case arises if the probability $\hat{p}_n$ is estimated to be zero, meaning the event defining the property did not occur. In this case a confidence interval can be defined by looking at the probability of the event not occurring in any of $n$ samples, which is $p_n = (1-p)^n$. Assuming that $p_n \geq 0.01$, and using a Taylor approximation for large $n$, a 99% confidence interval of $[0, 4.605/n]$ is a reasonable assumption. For 95% confidence intervals the region of $\hat{p}_n \pm 1.96\,\hat{\sigma}_n$ is taken, changing the factor in (6) from 2.58 to 1.96. In the same fashion, the 95% confidence interval for events that did not occur during the study is $[0, 3/n]$, where n is the number of samples [12].

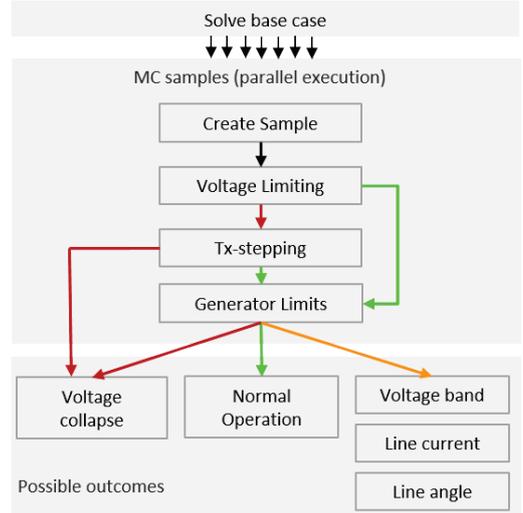

Figure 2 Algorithm of the SUGAR Monte Carlo simulations.

## D. Load Uncertainties

Power system operators rely on load forecasting to schedule generation ahead of time. One way of quantifying its accuracy is by using the confidence interval estimation. Reference [13] evaluates load forecast errors from 4 years of load forecasts of the French power grid to find confidence intervals for the forecast. It depicts different probability distribution functions (PDF) of forecast errors for different time horizons. One can see that for short term forecasting (one-hour intervals) an error PDF resembles a distribution close to a normal distribution, while for longer term forecasts (days, months) more complex PDFs are depicted. The case studies in this paper assume to deal with short term probabilistic contingency analyses during power system operation, which occurs for very short time periods such as the time interval from one state estimation to the next. Hence, normal distribution of load uncertainty is a good assumption for the cases studied in this paper. To find a quantification of this uncertainty for transmission networks it is assumed that the lowest level of a transmission network coincides with the highest level of a distribution system. A study on distribution system short-term load forecasting errors is presented in [14] using different forecast models on Irish and Danish smart meter data over two years for different distribution system levels. The study finds the load forecasting errors for the highest distribution system level to be in the order of a few percent. This paper will use load uncertainty values of 1% and 2% in the case studies presented in Section IV.

## IV. RESULTS

The following results were generated using the approach described in the previous section, which we refer to as the SUGAR MC algorithm. To provide a baseline, a comparison to Monte Carlo simulations using a standard power flow tool is presented. Successively, two power flow cases are studied using the SUGAR MC algorithm. A case study of a 145-bus test system showing high probability of angular instability will be presented, including a contingency study and a comparison of uniform and normal distributed loads. Finally, a probabilistic contingency analysis of the 13659-bus PEGASE test case will be presented.

### A. Comparison of the proposed aproach to a standard PF tool

The 13659-bus PEGASE system is used to compare the performance of the SUGAR MC algorithm to a standard power flow tool. For this study, the real power P, as well as the reactive power Q of each load in the system are assumed to be normally distributed with standard deviations $\sigma_{PQ}$ of 1% to 10% of nominal P and Q values. The SUGAR MC algorithm creates 1000 samples for each standard deviation value in the graph and solves the corresponding PF cases. The exact load values of each sample are stored and reused by the standard PF tool, enabling a one to one comparison. For both tools, the PF algorithm is initialized with the original system's solution. As a comparative measure, the ratio of failed samples to the total number of samples is taken as an estimate for the probability of grid collapse. Figure 3 shows the estimated probability of grid collapse for both tools over different standard deviations of normally distributed PQ loads as well as the 95% confidence intervals of those estimates calculated by the aforementioned method. For each simulated scenario, SUGAR finds the estimated probability of grid collapse to be lower than the estimation of the standard PF tool. On average, the estimated values of SUGAR are found to be 19% lower over the range of this study, illustrating that robustness of the simulation tool is a pre-requisite to make probabilistic statements about power systems.

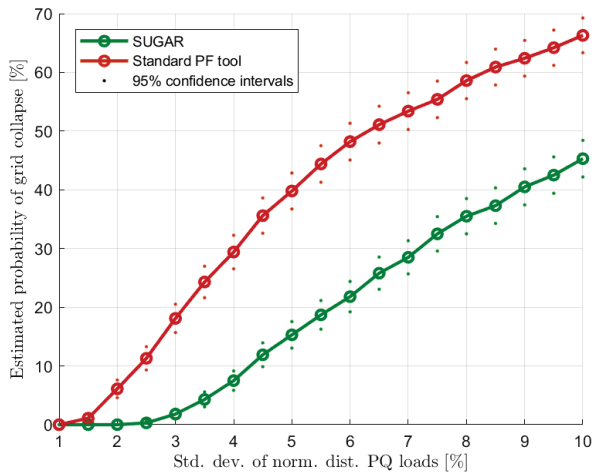

Figure 3 Comparison of estimated probability of grid collapse for PQ load uncertainties of the 13659 bus PEGASE system between SUGAR and a standard PF tool.

Even though the robustness of power flow analysis is greatly improved by SUGAR, there is no guarantee that a system for which no solution was found does not have a feasible solution in the actual system that is being modeled. However, the robust convergence to solutions for difficult power flow cases with our approach provides tightened bounds to physically correct scenarios, thereby enabling power system operations with less conservative margin. The results provided by this approach can still be seen as conservative, since the case of a false positive, namely, a numerical solution without a corresponding physical solution, is very unlikely if the modeling of the system is physically correct. For samples that do not converge using the SUGAR approach, further studies are possible to verify the infeasibility of the sample. For example, the trajectory of the Tx-stepping method could be studied to find buses that experience voltage collapse or lines that show angular instability during the gradual changes of the system. A similar approach can be taken by studying a continuation power flow [15] that attempts to find the point of voltage collapse by perturbing the loading factor of the system. Additionally, cases that did have a solution but did not converge after the generator reactive power limits are set can be studied to prove their infeasibility. Molzan, et.al. [16] provide a method to find sufficient conditions for this proof as well as a detailed overview of existing work towards proving power flow infeasibility.

### B. Case Study of a 145 bus system

A first case study will be presented on a publicly available 145-bus system. It will be shown that it has interesting properties such as a non-zero probability of voltage collapse and angular instability for moderate values of load uncertainty. First, the effect of different uncertainty distributions of P and Q loads will be studied, followed by a discussion of the reason for angular instability. Finally, a probabilistic contingency study of this case will be presented.

#### 1) Influence of probabilistic PQ distribution

To study the effect of different distributions for the 145-bus case study, two Monte Carlo simulations were run. The first for normally distributed loads, with a standard deviation of $\sigma_{PQ} = 1\%$ of the nominal P and Q values. The second study included uniformly distributed P and Q load values with a range of ±3% of the nominal values to represent the range of 99.87/% of values or 6 $\sigma_{PQ}$ of the first experiment.

Table 1: Estimated probabilities of voltage collapse and angular instability for uniformly as well as normally distributed real and reactive power loads.

| 145-bus case | Voltage Collapse ($\hat{p} \pm ci_{99}$) [%] | Angular instability ($\hat{p} \pm ci_{99}$) [%] |
|---|---|---|
| Uniform distr. $\Delta_{PQ}$± 3% | 14.28 ± 0.29 | 16.50 ± 0.30 |
| Normal distr. $\sigma_{PQ} = 1\%$ | 1.50 ± 0.10 | 10.85 ± 0.25 |

Table 1 shows the result of the experiment with uniform distributions. It finds an estimated probability of voltage collapse to be around 14.28%, and an estimated probability that at least one line hit the voltage angle limit of 16.5%, in a 99% confidence interval of 0.29% and 0.3%, respectively. For normally distributed PQ-loads these values are much lower, which is easily explained by the increased probability of extreme samples for the uniform distribution.

Table 2: Estimated probabilities of voltage collapse and angular instability for normally distributed P and Q loads for different operational states of the system. Such as normal operation and contingencies C1 to C3.

| $\sigma_{PQ} = 1\%$ Normal distribution | Voltage Collapse ($\hat{p} \pm ci_{99}$) [%] | Angular instability ($\hat{p} \pm ci_{99}$) [%] |
|---|---|---|
| Normal operation | 1.50 ± 0.10 | 10.85 ± 0.25 |
| C1: G132 | 1.69 ± 0.11 | 16.88 ± 0.31 |
| C2: B130-131 | 1.51 ± 0.09 | 7.03 ± 0.21 |
| C3: B130-131 B131-144 | 1.52 ± 0.09 | 98.48 ± 0.01 |

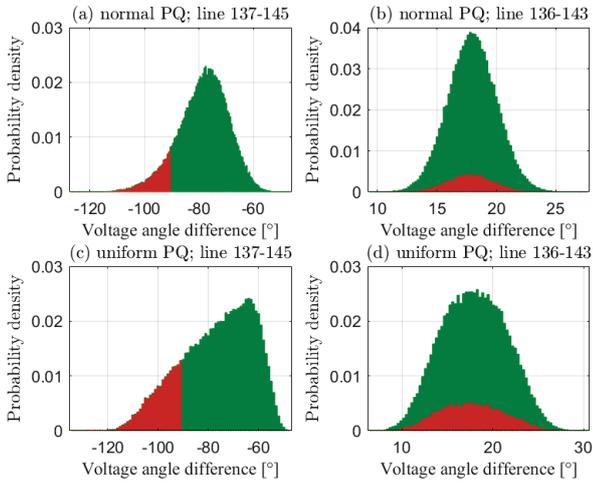

Figure 4 Probability density function of voltage angle differences for normally distributed P and Q in (a) for line 137-145 and in (b) for line 136-143 as well as uniformly distributed P and Q in (c) for line 137-145 and in (d) for line 136-143.

Figure 4 shows the resulting probability density functions (PDFs) of voltage angle differences for two lines in the system. The PDFs correspond to the PDF found by evaluating every feasible sample of the studies. The color code of the plots in Figure 4 corresponds to samples that are accepted as normal power system operation (green) and samples that were found to have at least one angular unstable line in the system (red). While plot (a) and (b) correspond to the study with normally distributed loads, plot (c) and (d) correspond to the study including uniformly distributed loads. The highly nonlinear relationship between P and Q values and the voltage angle difference is visible in plots (a) and (c) that show a skew to the right. Plots (b) and (d) resemble the underlying distributions of P and Q in a more linear fashion even though one can see in (d) that, per central limit theorem, multiple independent probabilistic influences generate a PDF that is converging to a normal distribution.

*2) Angular instability*

Referring to Figure 4, one can see in plots (a) and (c) that the samples for line 137-145 hit the maximum allowable voltage angle difference of 90° and are marked angular unstable, clearly indicating the reason for the angular instability of the system in the line from bus 137 to 145. Plots (b) and (d) show the same angular unstable samples marked in red, exposing the share of unstable sample values in (b) and (d). The accumulated PDF in (b) and (d) is created by the sum of the samples of normal operation and angular unstable operation. The sub distributions of PDFs in (b) and (d) show no visible correlation with angular instability.

*3) Probabilistic contingency analysis*

A final study on the 145-bus case is a probabilistic contingency analysis. Contingency analyses were run for N-1 and N-2 contingencies including N-1 outage scenarios for the five biggest generators, N-2 generator outage scenarios for the biggest 2 generators after the first outage. In addition, N-1 contingencies of the 10 highest loaded branches were simulated. All generator contingencies were found to have a feasible base case, in contrast to branch contingencies where only six out of 10 base cases were found to be feasible. In the current version of the SUGAR MC algorithm the program aborts when there is no base case solution, thus assigning a 100% estimated possibility of voltage collapse. However, as the Tx-stepping method does not rely on a good initial guess, a slight change in the algorithm would give the possibility of Monte Carlo simulations without a baseline solution. This would be enable the algorithm to find an estimated probability of system collapse for cases that do not have a feasible base line solution. Table 2 shows estimated voltage collapse and angular instability probabilities for the original 145-bus case and selected (interesting) contingencies. For contingency C1, a generator outage on bus 132, a rise in the estimated probability of voltage collapse as well as the estimated probability of angular instability over the base case is seen. Contingency C2, a line outage on between buses 130 and 131, shows only a slight increase in the estimated probability of voltage collapse, but it surprisingly shows a decrease in the probability of angular instability compared to the base case. Contingency C3, an additional line outage to C2 between buses 131 and 144, shows a slight rise in the probability of voltage collapse compared to C2, but brings angular instability with an estimated 100%. Every sample of this contingency that did not result in voltage collapse was found to be angular unstable. The select three contingencies demonstrate well how a probabilistic approach gives a much more detailed picture of the different risks different contingencies create.

*C. Contingency study of the 13659 bus PEGASE system*

To show the scalability of this approach, a probabilistic contingency analysis is conducted for the 13659-bus PEGASE power system case. For this study, the load uncertainty is assumed to be normally distributed with a standard deviation of $\sigma_{PQ} = 2\%$.

Table 3 shows the contingencies that were studied. First the N-1 contingencies of the 10 biggest generators are run, followed by the N-2 contingencies of the three biggest remaining generators. Finally, N-2 contingencies of the 10 biggest generators and the 2 highest loaded branches are examined. The contingency involving a generator outage on bus 7640 did not create feasible base cases. As already noted, a slight change in the algorithm could estimate the probability of voltage collapse for these cases as well.

Table 3 Feasibility of the Contingencies of the 13659-bus PEGASE system base cases.

| 13659-bus PEGASE contingencies | Infeasible | Feasible |
|---|---|---|
| N-1: 10 Biggest generators | 1 | 9 |
| N-2: 10x3 generators | 3 | 27 |
| N-2:10 generators x 2 branches | 2 | 18 |

To quantify the estimated probabilities of voltage collapse and angular instability for this system, the 13659-bus PEGASE base case and the select derived contingencies are studied more closely. Table 4 quantifies the probabilities of voltage collapse for the 13659-bus PEGASE system as well as select contingencies with nonzero estimated voltage collapse probabilities. The base system converged to a zero voltage collapse probability with a 99% confidence interval of [0.0 0.02 0.05]%, in 5159 simulations one sample was found infeasible. In a boundary case like this, where the original confidence interval slightly goes into negative probabilities, the lower boundary of the confidence interval is set to zero [12]. Contingency C1 is a generator outage on bus 2067 resulting in an estimated probability of voltage collapse of 1.46%. Contingency C2 is a two-generator outage scenario with an estimated probability of voltage collapse of 2.27%. Finally, contingency C3 is a N-2 contingency with the same generator as contingency C1, here an additional branch outage from bus 12854 to 6522 lowers the probability of voltage collapse compared to C1.

Table 4 Estimated Probabilities of voltage collapse and angular instability for the 13659-bus PEGASE system and derived contingency cases.

| $\sigma_{PQ} = 2\%$ Contingencies | Voltage Collapse ($\hat{p} \pm ci_{99}$) [%] | Angular instability ($\hat{p} \pm ci_{99}$) [%] |
|---|---|---|
| Normal operation | [0.0 0.02 0.06] | 0.00 + 0.05 |
| C1: G2067 | 1.46 ± 0.30 | [0.00 0.01 0.04] |
| C2: G12587;G11825 | 2.27 ± 0.38 | [0.00 0.07 0.14] |
| C3: G2067;B12854-6522 | 0.63 ± 0.20 | 0.00 + 0.05 |

V. Conclusion and future work

This paper presented a Simple Random Sampling Monte Carlo approach using SUGAR, a robust power flow approach based on the equivalent circuit formulation including continuation methods to reliably find the steady state operating point of large scale power systems. Based on a one-to-one comparison of Monte Carlo samples of the 13659-bus PEGASE case with a standard power flow tool, it was shown that a robust solver is a prerequisite for valid probabilistic statements on a power system. Furthermore, the 145-bus system and its property of a non-zero chance of angular instability as well as voltage collapse were studied, and the probabilistic contingency analyses were presented for it as well as the 13659-bus PEGASE system. The results demonstrated the increase in valuable information provided by a probabilistic approach to contingency studies. Future work will be including uncertainties of generation, studies for different uncertainties and distributions on select buses or for select equipment. New work will also include measurement data uncertainties, which is straight forward to incorporate using the equivalent circuit approach.

Acknowledgements

This work was supported in part by the Defense Advanced Research Projects Agency (DARPA) under award no. FA8750-17-1-0059 for the RADICS program.

VI. References

[1] H. W. Dommel, W. F. Tinney, and W. L. Powell, "Further Developments in Newton's Method for Power System Applications," in *IEEE Winter Power Meeting, New York, N.Y., 1970*, 1977.
[2] V. M. da Costa, N. Martins, and J. L. R. Pereira, "Developments in the newton raphson power flow formulation based on current injections," *IEEE Trans. Power Syst.*, vol. 14, no. 4, pp. 1320–1326, 1999.
[3] A. Pandey, M. Jereminov, G. Hug, and L. Pileggi, "Improving Power Flow Robustness via Circuit Simulation Methods," in *IEEE PES General Meeting*, 2017.
[4] A. Pandey, M. Jereminov, M. R. Wagner, and L. Pileggi, "Robust convergence of Power Flow using Tx Stepping Method with Equivalent Circuit Formulation (Accepted)," in *Spower Systems Computation Conference, Dublin*, 2018.
[5] W. Li, Y. Mansour, J. K. Korczynski, and B. J. Mills, "Application Of Transmission Reliability Assessment in Probabilistic Planning Of BC Hydro Vancouver South Metro System," *IEEE Trans. Power Syst.*, vol. 10, no. 2, pp. 964–970, 1995.
[6] A. M. Leite da Silva, J. F. C. Castro, and R. A. Gonzalez-Fernandez, "Spinning reserve assessment via quasi-sequential Monte Carlo simulation with renewable sources," in *2016 International Conference on Probabilistic Methods Applied to Power Systems (PMAPS)*, 2016, pp. 1–7.
[7] T. Cui and F. Franchetti, "A Quasi-Monte Carlo approach for radial distribution system probabilistic load flow," *2013 IEEE PES Innov. Smart Grid Technol. Conf.*, pp. 1–6, 2013.
[8] A. C. Melhorn, A. Dimitrovski, and A. Keane, "Probabilistic load flow: A business park analysis, utilizing real world meter data," *2016 Int. Conf. Probabilistic Methods Appl. to Power Syst. PMAPS 2016 - Proc.*, 2016.
[9] Z. Yang, "A Probabilistic Load Flow Method Considering Transmission Network Contingency," in *Power Engineering Society General Meeting*, 2007, pp. 1–6.
[10] Y. Chen, P. Etingov, H. Ren, Z. Hou, M. Rice, and Y. V. Makarov, "A look-ahead probabilistic contingency analysis framework incorporating smart sampling techniques," *IEEE Power Energy Soc. Gen. Meet.*, vol. 2016–Novem, no. Mc, 2016.
[11] D. M. Bromberg, M. Jereminov, X. Li, G. Hug, and L. Pileggi, "An Equivalent Circuit Formulation of the Power Flow Problem with Current and Voltage State Variables."
[12] A. Owen, *Monte Carlo theory, methods and examples*. 2013.
[13] B. Petiau, "Confidence interval estimation for short-term load forecasting," *Evolution (N. Y).*, pp. 1–6, 2009.
[14] B. Hayes, J. Gruber, and M. Prodanovic, "Short-Term Load Forecasting at the local level using smart meter data," in *2015 IEEE Eindhoven PowerTech*, 2015, pp. 1–6.
[15] V. Ajjarapu and C. Christy, "The continuation power flow: a tool for steady voltage stability analysis," *Ieee*, vol. 7, no. 1, pp. 304–311, 1992.
[16] D. K. Molzahn, V. Dawar, B. C. Lesieutre, and C. L. Demarco, "Sufficient conditions for power flow insolvability considering reactive power limited generators with applications to voltage stability margins," *Proc. IREP Symp. Bulk Power Syst. Dyn. Control - IX Optim. Secur. Control Emerg. Power Grid, IREP 2013*, 2013.